Exploring the Unconventional Electron Distribution Patterns in Iron-based Superconductors

[1,2]Chi Ho Wong*, [1]Rolf Lortz*

[1]Department of Physics, The Hong Kong University of Science and Technology, Hong Kong

[2]Department of Industrial and Systems Engineering, The Hong Kong Polytechnic University, Hong Kong

Emails: chkhwong@ust.hk, lortz@ust.hk

**Abstract:**

For more than a decade, the unusual distribution of electrons observed in ARPES (angle-resolved photoemission spectroscopy) data within the energy range of ~30meV to ~300meV below the Fermi level, known as the ARPES range, has remained a puzzle in the field of iron-based superconductivity. However, in this study, we have made a noteworthy observation: although the electron-phonon coupling alone is insufficient to account for the observed ARPES pattern, our analysis reveals that when the instantaneous electron-phonon coupling occurring in selective phonon channel is enhanced by the coexistence of antiferromagnetic spin density wave and charge density wave phenomena, the amplified interaction becomes comparable to the ARPES range. This finding suggests that the instantaneous interplay between these intricate phenomena should play a crucial role in generating the observed energy range in ARPES. Our work may provide a valuable clue towards achieving a deeper understanding of the complex relationship between electronic correlations, lattice structure, and superconductivity in iron-based materials for uncovering the origin of the unconventional ARPES pattern.

**Introduction:**

Iron-based superconductors have attracted considerable attention in the field of condensed matter physics due to their unique properties and potential applications in high-temperature superconductivity [1,2,3]. However, there are numerous complex features that still lack a comprehensive understanding, such as the intricate effects of antiferromagnetism, nematicity, gap anisotropy, spin-orbit coupling, and other factors on the underlying pairing mechanism [1,2,3]. In particular, the experimental ARPES studies have confirmed the unusual electron distribution pattern below the Fermi level, where the electrons within the energy range of ~30meV to ~300meV below the Fermi level can be surprisingly affected by iron-based superconductivity and this unusual pattern has remained a longstanding puzzle in the study of iron-based superconductors [4-6]. In spite of extensive research efforts over the past decade, the origin of this unconventional electron distribution pattern has remained elusive. Understanding the science underlying the unusual energy range observed in ARPES is crucial as it may provide insights into the triggers of iron-based superconductivity.

To unravel the origin of the unconventional electron distribution pattern in the ARPES data [4-6], our analysis goes beyond the conventional electron-phonon coupling model and explores the interplay between various electronic correlations, lattice structure, and key phenomena such as antiferromagnetic spin density wave (SDW) and charge density wave (CDW) in the tetrahedral regions. The presence of SDW can provide a two-fold increase in the antiferromagnetically enhanced electron-phonon scattering matrix [7]. Apart from this, the SDW can induce the electric xy potential in the form of the CDW phenomenon, and further increase the electron-phonon scattering matrix by another factor of ~2 [7]. These two amplification factors (Coh factor = $R_{SDW} \cdot R_{CDW}$) of electron-phonon coupling were proposed by Coh et al [7], in which the non-cancellable out-of-plane phonon in the tetrahedral zones between the magnetic and non-magnetic lattice sites under antiferromagnetic spin density wave was observed under enormous experimental and computational costs, resulting in a significant improvement in the agreement between simulation results and experimental observations [7,8]. By employing their proposed method [7,8], it becomes possible to explicitly demonstrate the emergence of an induced xy potential in the presence of complex phenomena. Despite this finding providing a more accurate calculation of the experimental behavior observed in the iron-based superconductors, the fundamental understanding of the amplification effects may not be comprehensive.

In this paper, we delve into the fundamental understanding of the Coh factors [7] and investigate whether antiferromagnetism, spin density wave, and charge density wave are the major ingredients in the ARPES range. By examining their interplay, we aim to unmask the complex mechanisms that drive the unconventional electron distributions in the ARPES data [4-6]. Such understanding is crucial for advancing the knowledge of these iron-based compounds and facilitating the development of new strategies for the theory of iron-based superconductors.

**Computational Methods**

For a lattice ion located at position $R_i$, with a displacement $u_i(x, y, z)$ from its equilibrium position $R_i^0$, and assume that the potential $V(r)$ of the ion is rigid, then the electron-phonon interaction can be expressed as $H = \sum_{i\sigma} \int \psi_\sigma^\dagger(r)\psi_\sigma(r)V(r-R_i)d^3r$ where $\psi$ is the wavefunction of electrons and $\sigma$ is spin index. For small amplitude vibration, it can be expanded as [9]

$$H = \sum_{i\sigma} \int d^3r \psi_\sigma^\dagger(r)\psi_\sigma(r)V(r-R_i^0) + \sum_{i\sigma} \int d^3r \psi_\sigma^\dagger(r)\psi_\sigma(r) u_i \cdot \nabla_{R_i} V(r-R_i^0)|_{R_i^0} + ...$$

Antiferromagnetism can enhance electron phonon scattering matrix by a factor of $R_{AF}$ [10]. In the presence of spin-density wave, the antiferromagnetic fluctuations are expected to form AFM local minima and maxima alternatingly [7]. While the AFM fluctuations in the local minima are rare, the spin density wave maximizes the AFM fluctuations at the local maxima that create a twofold increase in the local electron-phonon scattering matrix at the maxima. Meanwhile, the non-cancellable or differential vibrational amplitude along the out-of-plane axis (i.e. $\delta u_i(z) = |u_i(z_{peak})| - |u_{i+1}(z_{peak})| \neq 0$) of the nearest Fe neighbors occurs and then induce an additional electrostatic potential along the xy plane in the form of CDW [7], or equivalently amends the electronic density of states in the xy plane, which further increases the electron-phonon scattering matrix by a factor of $R_{CDW}$ [7]. It is important to note that the square of the electron-phonon scattering matrix is proportional to the electron-phonon coupling strength [10].

The AFM & SDW amplified electron-phonon interactions on the Fermi level are estimated for the compressed bulk FeSe and 2D FeSe/SrTiO3. We apply CASTEP to conduct the spin-unrestricted calculations at the GGA-PW91 level [11,12] for estimating the electronic density of states, band diagram, exchange correlation energy $E_{co}$ and magnetic moment $M_{Fe}$. The influence of exchange coupling to the antiferromagnetically assisted electron phonon coupling can be rewritten through the separation of variable, in which the electron phonon coupling is multiplied by the exchange enhancement factor [13]. The exchange enhancement factor in the form of Ising expression $f(E_{ex}) \sim \dfrac{M_{Fe}^2 E_{co}|_{P>0}}{M_{Fe}^2 E_{co}|_{P=0}}$ is used to monitor the AFM & SDW amplified electron-phonon coupling under pressure P, where $R_{AF}^2|_{P>0} \sim R_{AF}^2|_{P=0} \cdot f(E_{ex})$ and $R_{tetra}^2|_{P>0} \sim R_{tetra}^2\|_{P=0} \cdot f(E_{ex})$ [13].

The maximum SCF cycle is set to 100 with a tolerance of 2 μeV/atom and the reciprocal k-space grid is 0.025(1/Å). The ultrasoft pseudopotential is used. The phonon data is calculated by the finite displacement mode, where the cutoff radius is 0.5nm and the interval of the dispersion is 0.04(1/Å). The non-cancellable or differential vibrational amplitude along the out-of-plane axis is interpreted using a ground-state harmonic oscillator model [10]. To evaluate the relevance and significance of AFM and SDW effects in the observed ARPES pattern [4-6], the AFM & SDW amplified electron-phonon couplings are calculated in physical units.

**Results and Discussions:**

After emerging the non-cancellable out-of-plane phonon [7] under the influence of the antiferromagnetic spin density wave, we observe that the average phonon frequency of bulk FeSe is decreased by ~2%. This decrease is accompanied by a relative out-of-plane vibrational amplitude between the two adjacent Fe atoms, which is approximately 0.04Å. When compared to the norm of $u_i(x, y, z)$ in bulk FeSe, typically ~0.15Å, the differential lattice vibrations $\delta u_i(z)$ are significant. The differential orthogonal displacement of lattice ions is observed not only in the FeSe system but also in the FeAs system. For example, $Ba_{0.6}K_{0.4}Fe_2As_2$ superconductor [14], where their relative out-of-plane lattice vibrations are ~0.07Å. These findings suggest that the presence of alternating AFM and SDW ordering triggers the abnormal out-of-plane lattice vibrations in the iron-based superconductors. In particular, the magnitude of $\delta u_i(z)$ in FeSe is smaller compared to that in the K-doped $BaFe_2As_2$. This is attributed to the fact that the atomic spring constant of the FeSe bond is approximately two times weaker than that of the FeAs bond where a weak atomic spring constant is less effective to create an orthogonal phonon in the tetrahedral zone.

The 11-type iron-based superconductor, bulk FeSe, exhibits a superconducting transition temperature ($T_c$) of approximately 10K under ambient pressure [15]. Notably, the $T_c$ of bulk FeSe can be enhanced by applying compression [15]. Our calculations indicate that the bare electron-phonon coupling in bulk FeSe, assuming isotropic momentum space, is determined to be 1.8 meV at 0GPa. It is important to highlight that the bare electron-phonon coupling strength exhibits a decreasing trend with increasing pressure, as depicted in Figure 1a. This observation indicates that the electron-phonon coupling alone is insufficient to explain both the observed ARPES range and the dependence of $T_c$ on pressure. However, the magnetic moment of the Fe atoms and the exchange-correlation energy increases with pressure, which convinces us to combine antiferromagnetism with electron-phonon coupling to check if the antiferromagnetically assisted electron-phonon coupling can produce such strong ARPES energy range. After activating the spin-unrestricted mode in the ab-initio calculation, the antiferromagnetically assisted electron-phonon coupling of bulk FeSe raises to 2.7meV only, where $R_{AF} \sim 1.2$. However, when it exhibits the coexistence in AFM and SDW interactions, the redistribution of AFM fluctuations between the local AFM minima and local AFM maxima contributes a 4-fold boost in the electron-phonon interaction (10.8meV). Furthermore, the effect of the out-of-plane phonon strengthens the interaction to 47.6 meV, where $R_{CDW}$ is justified to be 2.1 in FeSe [7]. As an anisotropic momentum space is observed in FeSe [16], the AFM & SDW electron-phonon interaction is then paled to 27.6 meV, which is comparable to the experimental ARPES range [4,6]. In unconventional superconductors, a reduction in the electron-phonon coupling strength has been observed due to the presence of symmetry effects in momentum space. Specifically, when a 4-fold symmetry is present in the superconducting gap, the electron-phonon coupling can be diminished by a factor of approximately 0.6-0.8 [17]. Our analysis reveals that the reduction in the interaction attributed to the gap anisotropy is approximately 0.58 in bulk FeSe, which is consistent with previous findings reported in the literature.

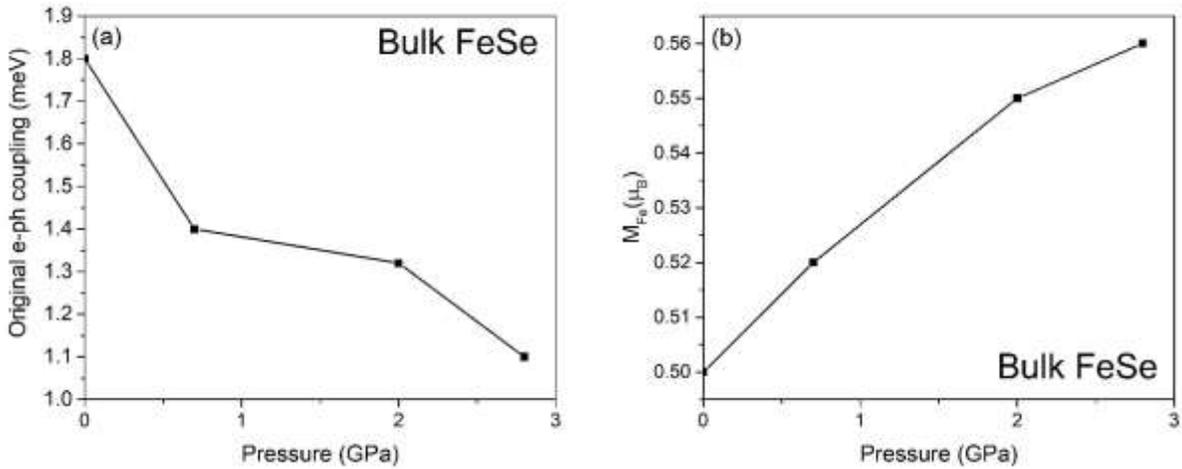

Figure 1: (a) The bare electron-phonon coupling of FeSe. (b) The magnetic moment of Fe atoms increases with pressure, where the exchange-correlation energy increases by a few percent upon compression.

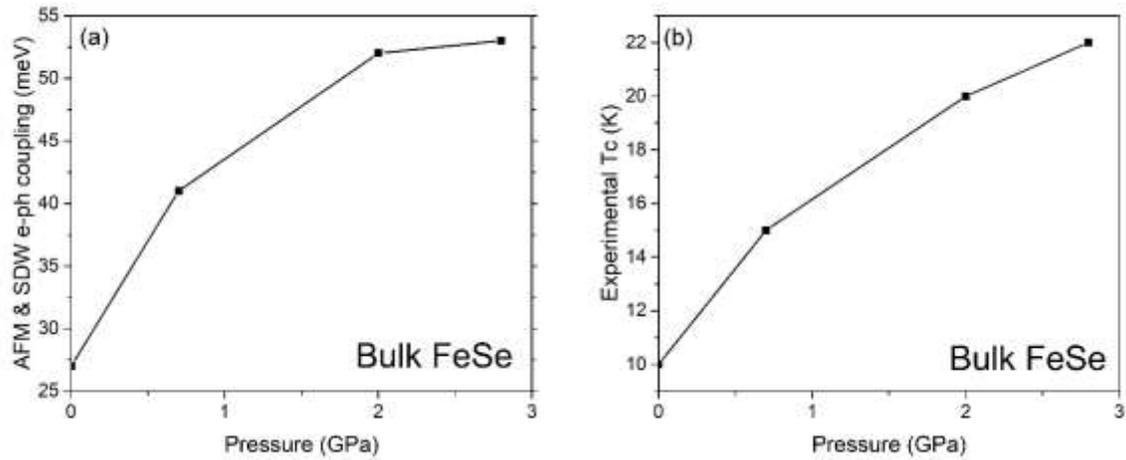

Figure 2: Bulk FeSe; (a) AFM & SDW enhanced electron-phonon coupling. (b) The $T_c$ measured in the experiments.

To validate the comparability between the AFM & SDW amplified electron-phonon coupling and the ARPES range, it is essential to select another iron-based compound that exhibits a substantial contrast in the ARPES range. Therefore, our upcoming investigation concentrates on the FeSe/SrTiO$_3$ composite with a high $T_c$ of ~100K in which its ARPES range in the experiment covers a broader energy range of ~0.1-0.3 eV [5,18-20], allowing for a comprehensive comparison and analysis. Following the same approach, our simulations have revealed that the influence of the differential out-of-plane phonon on the AFM & SDW electron-phonon scattering matrix in the FeSe/SrTiO$_3$ is approximately 1.3 times greater than in bulk FeSe. This difference arises from the structural characteristics of the FeSe/SrTiO$_3$ interface. In FeSe/SrTiO$_3$, the upper tetrahedral region of FeSe film is exposed to a vacuum environment, while the lower tetrahedral region interacts with the SrTiO$_3$ substrate. This differential proximity interaction further induces an imbalance in the out-of-plane lattice vibrations, leading to a more pronounced effect on the differential out-of-plane phonon compared to the bulk FeSe case. The electron-phonon coupling of FeSe/SrTiO$_3$ reinforced by the antiferromagnetic spin density wave is estimated to be ~0.5 eV, falling within the same order of magnitude as the experimental ARPES range. Notably, the experimental observation also demonstrates a significant strength in the electron-phonon coupling in FeSe/SrTiO$_3$, with the interfacial phonon reaching as high as 1150K [19], where its ultra-strong electron-phonon coupling has been reported [18-20]. By drawing a parallel between bulk FeSe and FeSe/SrTiO$_3$, we observe a proportional relationship between the AFM & SDW amplified electron-phonon interaction and the ARPES range.

The results obtained from the iron-based compounds raise a valid concern that the antiferromagnetic effect on the electron-phonon coupling may have been significantly underestimated. While the antiferromagnetically assisted electron-phonon coupling alone may not exert a significant impact on the electrons below the Fermi level, the interplay between antiferromagnetism and spin density wave synergistically enhances the electron-phonon interaction to a magnitude comparable to the unusual energy range observed in ARPES data. This amplification may demonstrate the crucial role played by the combined influence of antiferromagnetism and spin density wave effects in shaping the electron-phonon interactions. Consequently, the lattice ions may exert a much stronger electrostatic attraction, influencing the behavior of electrons within the ARPES range during the formation of Cooper pairs.

Calibrating the numerical parameters to achieve a good agreement with experimental observations can be an important step toward comprehending the underlying phenomena [7,8]. The subsequent step should involve delving into the scientific principles that elucidate why such agreement is achieved. This deeper exploration aims to uncover the fundamental mechanisms and processes that contribute to the observed phenomenon and provide a robust scientific understanding of the phenomenon in question. The enormous increase in electron-phonon coupling observed in iron-based superconductors can be understood through the physics behind the redistribution of spatial antiferromagnetic fluctuations under the effect of spin-density wave. This redistribution leads to the formation of alternating regions of AFM local minima and AFM local maxima between nearest neighbors of Fe atoms, in which the effect of spin density wave transfers the antiferromagnetic energy from the AFM local minima to the AFM local maxima per repeating unit. The SDW-induced redistribution of AFM fluctuations maintains the conservation of antiferromagnetic energy, but this energy transfer doubles the local electron-phonon scattering matrix at the AFM local maxima within the repeating unit. The process of AFM redistribution continues over time, periodically switching the locations of AFM local maxima and local minima, resembling an interference-like phenomenon. For instance, at time t, the odd-numbered lattice points represent the AFM maxima, while the even-numbered lattice points correspond to the AFM minima. However, as time progresses to half SDW period of oscillation, the AFM minima configuration switch to occur in the odd-numbered lattice points, while the AFM maxima configuration appears in the even-numbered lattice points. In other words, the alternating (odd and even) lattice points that form the AFM maxima and minima undergo changes in their configuration periodically over time. This interference-like effect significantly increases the peak value of the AFM & SDW enhanced electron-phonon coupling.

Furthermore, antiferromagnetism always slows down phonons in the lattice [13]. Lattice ions experiencing the AFM local maxima always vibrate slower, equivalently exhibiting a larger effective atomic mass. On the other hand, the neighboring lattice ions located at the AFM local minima vibrate relatively faster, indicating a smaller effective atomic mass. This differential effective mass among neighboring Fe atoms leads to the variation in their vibrational amplitudes orthogonally [7,10], triggering the non-cancellable out-of-plane phonons per repeating unit at the boundary between the non-magnetic and magnetic lattice sites. At this instantaneous time, based on Maxwell's equation, electrons experience an electric potential at the boundary, which generates the charge density wave (CDW) phenomenon, further enhancing the electron phonon coupling in the system. While AFM fluctuation persists, the instantaneously amplified electron-phonon coupling at the boundary occurs very often. Altogether, the combined effects contribute to the significant increase in the electron-phonon coupling in iron-based superconductors. In reality, AFM is not a mean field in iron-based superconductors. Hence, conventional DFT approach based on mean-field approximation may not be able to capture this instantaneous effects.

The Ising expression [10] is believed to capture the overall behavior of how antiferromagnetism changes under pressure that can make our simulation less dependence on the choice of DFT functional. Although the Ising expression may not provide exact quantitative values, it can still generate a general trend that offers valuable insights into the effects of pressure on antiferromagnetic behavior. Nonetheless, the primary objective of our study is to analyze whether the AFM and SDW amplified electron-phonon coupling, as well as the experimental ARPES range, exhibit the same order of magnitude. In this context, the accuracy of our work is already sufficient to validate these observations.

Our work indicates that the AFM & SDW amplified electron-phonon coupling not only matches the magnitude of the experimental ARPES range [4-6] but also follows the trend of the $T_c$ if iron-based superconductivity is detected. In our investigation of the impact of spin-orbital coupling on the enhanced electron-phonon coupling in the antiferromagnetic spin-density wave system [21], we observe that regardless of whether we include the spin-orbital coupling in our calculations or not, the changes observed in the AFM & SDW reinforced electron-phonon couplings of these samples are only around 10%. Hence, we propose that the primary factors influencing the observed ARPES range should be related to the AFM, SDW and CDW states. However, it is unfair if we underestimate the effect of spin-orbital coupling on the pairing mechanism of iron-based superconductors. The reason is that although the AFM & SDW amplified electron-phonon coupling and the ARPES range may be correlated, it is still an open question whether spin-orbital coupling is one of the ingredients to pull a trigger on activating iron-based superconductivity [21].

**Conclusions**

Our research findings suggest that the significantly amplified electron-phonon coupling, which arises from the presence of AFM and SDW, not only matches the experimental magnitude of the ARPES range but also exhibits a correlation with the superconducting transition temperature. These observations suggest that the instantaneous interplay between AFM, SDW, and additional CDW states should play a critical role in shaping the ARPES range. The effects of the SDW-assisted AFM & CDW states have been analyzed, in which SDW triggers the redistribution of AFM fluctuation and then create CDW state. The presence of AFM, SDW, and CDW states in iron-based superconductors offers valuable insights into the fundamental mechanisms that drive the unconventional ARPES patterns. These insights may play a pivotal role in advancing the theoretical understanding of iron-based superconductors.